\documentclass[pra,twocolumn,aps,superscriptaddress,longbibliography]{revtex4-1}

\usepackage{amsmath}
\usepackage{amssymb}
\usepackage{graphicx}
\usepackage[usenames,dvipsnames]{color}
\usepackage{braket}
\usepackage{bm}
\usepackage{times}
\usepackage{hyperref}
\usepackage{pdfpages}
\usepackage{float}
\usepackage{lipsum}
\usepackage[dvipsnames]{xcolor}
\hypersetup{
  colorlinks=true,
  citecolor=blue,
  linkcolor=blue,
  urlcolor=blue}

\makeatletter
 \AtBeginDocument{\let\LS@rot\@undefined}
\makeatother

\begin{document}

\title{Quantum Hall states for $\alpha = 1/3$  in optical lattices}

\author{Rukmani Bai}
\email{rukmani20891@gmail.com}
\affiliation{Physical Research Laboratory,
             Ahmedabad - 380009, Gujarat,
             India}
\affiliation{Indian Institute of Technology Gandhinagar,
             Palaj, Gandhinagar - 382355, Gujarat,
             India}
\author{Soumik Bandyopadhyay}
\email{soumik@prl.res.in}
\affiliation{Physical Research Laboratory,
             Ahmedabad - 380009, Gujarat,
             India}
\affiliation{Indian Institute of Technology Gandhinagar,
             Palaj, Gandhinagar - 382355, Gujarat,
             India}
\author{Sukla Pal}
\email{sukla.ph10@gmail.com}
\affiliation{Physical Research Laboratory,
             Ahmedabad - 380009, Gujarat,
             India}
\author{K. Suthar}
\email{kuldeepphysics88@gmail.com }
\affiliation{Physical Research Laboratory,
             Ahmedabad - 380009, Gujarat,
             India}
\author{D. Angom}
\email{angom@prl.res.in}
\affiliation{Physical Research Laboratory,
             Ahmedabad - 380009, Gujarat,
             India}
%
\begin{abstract}
 We examine the quantum Hall (QH) states of the optical lattices with square 
geometry using Bose-Hubbard model (BHM) in presence of artificial gauge field. 
In particular, we focus on the QH states for the flux value of $\alpha = 1/3$. 
For this, we use cluster Gutzwiller mean-field (CGMF) theory with 
cluster sizes of $3\times 2$ and $3\times 3$. We obtain QH states at 
fillings $\nu = 1/2, 1, 3/2, 2, 5/2$ with the cluster size $3\times 2$ and
$\nu = 1/3, 2/3, 1, 4/3, 5/3, 2, 7/3, 8/3$ with  $3\times 3$ cluster.
Our results show that the geometry of the QH states are sensitive to the 
cluster sizes. For all the values of $\nu$, the competing superfluid (SF) 
state is the ground state and QH state is the metastable state. 
\end{abstract}



\maketitle
\section{Introduction}

  Ultracold Bosons in optical lattices (OLs) \cite{bloch_08} have been the 
subject of intense research since the experimental realization of Bose 
Einstein condensate (BEC) in OLs
\cite{anderson_98, greiner_01, greiner_02,lewenstein_07}.
The rapid developments in the control of BECs in OLs have made 
these systems an elegant experimental tool to explore the foundations of 
strongly correlated quantum many-body systems. Till recently, many of these 
were limited to the realm of theoretical studies. The near ideal, defect free, 
experimental realizations of OLs make these excellent proxies to explore
quantum many-body effects in condensed matter systems. One remarkable recent 
experimental development is the introduction of synthetic magnetic fields in 
OLs\cite{aidelsburger_13, miyake_13}. This makes the physics of 
quantum Hall (QH) effect accessible to OLs. An important outcome is the 
study of Harper-Hofstadter model \cite{harper_55,hofstadter_76} and observation
of fractal spectrum \cite{dean_13} for interacting Bosons in OL with synthetic 
magnetic field \cite{jaksch_03}. The scope to investigate the interplay of 
lattice geometry, synthetic magnetic fields, and strong interactions have made 
these systems an excellent platform to explore exotic quantum many-body phases. 

In the quantum description, the energies of electrons in an external magnetic 
field are quantized in to Landau levels. These have large degeneracy and in
a lattice these correspond to the Bloch bands \cite{harper_55} and is 
sensitive to applied magnetic field. The key point is the geometrical phase
an electron acquires when completing a loop in the cyclotron motion. Thus, 
neutral atoms in OLs can mimic the physics of electrons in magnetic fields
if a geometrical phase can be induced to the atoms. This is achieved through
the generation of a synthetic magnetic field in OLs through an artificial gauge
potential \cite{dalibard_11, lin_11, lin_09a, garcia_12} using lasers. Then,
an atom hopping around a single unit cell in the OL, also called a plaquette, 
acquires Peierls' phase \cite{peierls_33} of $\Phi = 2\pi\alpha$. Where 
$\alpha$ is the flux quanta per plaquette and it is related to the strength of 
the synthetic magnetic field. In the condensed matter systems realizing
high $\alpha$ require magnetic fields $\approx 10^3$ Tesla. 
In this respect, the OLs have the advantage that by suitable choice of 
external as well as internal parameters, various topological states such as 
fractional QH (FQH) states are obtainable with the current experimental 
realizations \cite{palmer_06, sorensen_05} and a variety of FQH-like states 
can be expected to emerge from these systems.   

A paradigmatic model which describes BECs in OLs is the Bose Hubbard 
model (BHM) \cite{fisher_89,jaksch_98}. In this model the kinetic energy of 
the bosons competes with the on site interaction and drives a quantum phase 
transition (QPT) from  superfluid (SF) to bosonic Mott insulator 
(MI) phase \cite{greiner_02, stoferle_04}. Various theoretical methods, such 
as mean field theory \cite{fisher_89}, strong coupling expansion 
\cite{freericks_96, niemeyer_99, freericks_09, wang_18}, quantum Monte Carlo 
\cite{wessel_04}, density matrix renormalization group \cite{peotta_14} have 
been used to study the role of quantum fluctuation and short range on site 
interaction on QPT. The SF phase is compressible with finite SF order parameter 
and phase coherent; MI phase on the other hand is incompressible 
with zero order parameter and shows integer commensurate filling per 
lattice site. In contrast to these two phases, the QH states are 
incompressible states with zero order parameter and have incommensurate 
filling. Several previous works 
\cite{sorensen_05,palmer_06,hafezi_07,umucalilar_07, palmer_08,umucalilar_10,
natu_16, hugel_17,kuno_17,gerster_17,bai_18} have theoretically explored the 
existence of FQH states in OLs using BHM with synthetic magnetic fields, the 
bosonic counterpart of the Harper-Hofstadter model
\cite{harper_55,hofstadter_76}. These theoretical works have also examined the 
possible signatures of the FQH states. One of the possibilities is the 
measurement of two-point correlation function in the bulk and in the edge of 
the lattice \cite{he_17}. Such measurements may be experimentally possible 
using the concepts from quantum information theory \cite{elliott_16,streif_16}.  

In the present work we use cluster Gutzwiller mean field (CGMF) theory to 
provide a better description of the atom-atom correlations. It is proven to 
be more accurate than single site Gutzwiller theory and previous works
have also used CGMF theory to study both the integer quantum Hall (IQH) and 
FQH states. In ref \cite{natu_16} the appearance of incompressible QH ground 
state in the hard-core limit with stripe order for $\alpha$ = 1/5 and
$\nu = 1/2$ is reported. Similarly, using reciprocal cluster mean field (RCMF) 
analysis, a competing FQH state is reported for $\alpha = 1/4$ in the recent 
study by H$\ddot{o}$gel {\it et al.} \cite{hugel_17}. Motivated by the above 
observation of QH states, we explore the possible QH states for $\alpha = 1/3$ 
with different cluster size in the hard core limit and demonstrate the 
dependence of the QH state geometry on the cluster size. For the present
studies we have considered $3 \times 2$ and $3 \times 3$ clusters and 
report the improvement in the description of the QH states with the 
increase of cluster sizes. This stems from more accurate accounting of the
correlation effect with larger cluster sizes. We have also performed a 
comparative study between the obtained QH and SF states with both the cluster
sizes and detect the emergence of various patterns.


\section{Theory}

We study a system of spinless bosonic atoms at T$ = 0$K, confined
in a two-dimensional (2D) optical lattice of square geometry under the 
influence of artificial gauge field 
\cite{jaksch_03,aidelsburger_11,aidelsburger_13,miyake_13}. 
In the Landau gauge $\mathbf{A} = (A_x, 0, 0)$ with $A_x = 2\pi\alpha q$, the 
system is described by the following Hamiltonian
~\cite{jaksch_98,jaksch_03,sorensen_05,palmer_06,palmer_08},
BHM Hamiltonian where Peierls substitution is incorporated in the 
nearest-neighbor (NN) hopping ~\cite{peierls_33,harper_55,hofstadter_76},
\begin{equation}
 \hat{H} = -\sum_{\langle jk\rangle}\left({\rm e}^{i2\pi\alpha q}J_x 
           + J_y \right ) \hat{b}_j^{\dagger}\hat{b}_k
           + \sum_j\hat{n}_j\left[\frac{U}{2}(\hat{n}_j-1) 
           - \mu )\right],
\label{bhm}         
\end{equation}
$j \equiv (p,q)$ corresponds to the lattice site index where $q$ is the index 
of the lattice site along $y$-axis, $\hat{b}_{j}$ ( $\hat{b}^{\dagger}_{j}$)
are the bosonic annihilation (creation) operators, $\hat{n}_{j}$ is the 
occupation number operator at $j^{th}$ lattice site, $J_x$ ( $J_y$) are the 
hopping strengths between two neighbouring sites along $x$ ($y$) direction, 
$U$ corresponds to the on-site interaction and $\mu$ is the chemical potential. 
Based on the experimental realizations, we consider isotropic hopping  
$J_x = J_y = J$, and repulsive on-site interaction energy ($U>0$). 
In presence of synthetic magnetic field, the atoms acquire a 
phase $2\pi\alpha$ upon hopping around a plaquette, where, $\alpha$ is 
the number of flux quanta per plaquette, and it has values
$0 \leq\alpha\leq 1/2$. In the absence of synthetic magnetic field 
($\alpha$ = 0), the Hamiltonian (\ref{bhm}) reduces to the familiar 
BHM Hamiltonian which admits two possible phases --- 
MI and SF~\cite{fisher_89,jaksch_98,greiner_02}. The MI phase appears in the
strongly interacting regime $(J/U\ll 1)$, while SF phase occurs in the 
limit $(J/U\gg 1)$. In homogeneous system, where the OL does not include any 
background potential, the phase-boundary between MI and SF forms lobes of 
different fillings and in presence of magnetic field the MI-lobes 
are enhanced \cite{oktel_07}. We employ the single site Gutzwiller mean-field 
method (SGMF) and CGMF to analyze the system in presence of synthetic magnetic
field and obtain the QH states.

%

\subsection{Mean Field Theory And Gutzwiller Approximation}

  Following the mean-field \cite{sheshadri_93} calculations of 
the BHM, we decompose the creation and annihilation operators 
in Eq.~(\ref{bhm}) into mean field and fluctuation around the 
mean-field, that is, $\hat{b}_{j} = \phi_{j} + \delta \hat{b}_{j}$, 
with $\phi_{j} = \langle\hat{b}_{j}\rangle$, and similarly,
$\hat{b}^{\dagger}_{j} = \phi^{*}_{j} + \delta \hat{b}^{\dagger}_{j}$, 
with $\phi^{*}_{j} = \langle\hat{b}^{\dagger}_{j}\rangle$. Neglecting the
term quadratic in the fluctuations, the mean-field Hamiltonian is
\begin{eqnarray}
\hat{H}^{\rm MF} &=&- \sum_{\langle jk\rangle} 
                  \left[(J_x {\rm e}^{i2\pi\alpha q} + J_y)
                  \left(\hat{b}_{j}^{\dagger}\phi_{k} 
                  + \phi^{*}_{j}\hat{b}_{k}
                  -\phi^{*}_{j}\phi_{k}\right) 
                  \right. \nonumber \\ 
                  && + \left.  {\rm h.c.}\right] 
                  + \sum_{j} \left[\frac{U}{2}\hat{n}_{j}
                  (\hat{n}_{j}-1) 
                  - \mu\hat{n}_{j}\right]. 
\label{mf_hamil}
\end{eqnarray}
We can, therefore, express the total Hamiltonian as the sum of single site
mean field Hamiltonians 
\begin{eqnarray}
\hat{h}_{j} =&-&\left [(J_x{\rm e}^{i2\pi\alpha q} + J_y) 
                  \left(\phi_{k}^*\hat{b}_{j} - \phi^{*}_{k}\phi_{j} \right) 
                  + {\rm H.c.}\right ]\\ \nonumber
                  &+& \frac{U}{2}\hat{n}_{j}(\hat{n}_{j}-1) 
                  - \mu \hat{n}_{j}.
   \label{ham_ss}
\end{eqnarray}
The next step is to diagonalize the Hamiltonian (\ref{mf_hamil}) for each 
site separately. For this, we consider the Gutzwiller ansatz, that is the 
ground  state of the entire lattice is the direct product of the ground 
states of all the individual sites, and can be written in the Fock basis as
\begin{eqnarray}
 |\Psi_{\rm GW}\rangle = \prod_j|\psi\rangle_j
                       = \prod_j \sum_{n = 0}^{N_{\rm b}}c^{(j)}_n
                         |n\rangle_j,
 \label{gw_state}
 \end{eqnarray}
with the normalization condition $\sum_{n} |c^{(j)}_n|^2$ = 1.
Here, $N_b$ is the occupation number state maximum number of particles at a
site and  $c^{(j)}_n$ corresponds the complex co-efficients for the 
ground state $|\psi\rangle_{j}$ at the $j$th site. Gutzwiller ansatz is 
the exact solution of the system in the strongly interacting regime ($J\ll U$). 

For the numerical computations, we consider $N_{\rm b} = 10$ and choose an 
initial guess of $\phi$. Then, we diagonalize the Hamiltonian for each site
and retain the ground state as the state $\ket{\psi}_j$ in 
$|\Psi_{\rm GW}\rangle$. Then, using this $|\psi\rangle_j$, we 
calculate new $\phi$ for the next iteration and this cycle is 
continued till convergence is reached. To distinguish the different phases, we 
compute the SF order parameter at each site, and  for the $j$th lattice site
SF order parameter is
\begin{equation}
\phi_j = \langle\Psi_{\rm GW}|\hat{b}_j|\Psi_{\rm GW}\rangle 
            = \sum_{n = 1}^{N_{\rm b}}\sqrt{n} 
              {c^{*(j)}_{n-1}}c^{(j)}_{n}.
\label{gw_phi}              
\end{equation}
From the above expression it is evident that $\phi_{j}$ is zero in the MI 
phase of the system since only one of the co-efficients in Eq.~(\ref{gw_state})
is non-zero, and it is finite for the SF phase due to the different $c_n$s 
contribution. We also compute the average lattice occupancy or density at 
each of the lattice site as    
\begin{equation}
\rho_j = \langle\Psi_{\rm GW}|\hat{n}_j|\Psi_{\rm GW}\rangle 
            = \sum_{n = 0}^{N_{\rm b}}n |c^{(j)}_n|^2.
\label{aver_n}              
\end{equation}
These are the essence of SGMF theory.


\subsection{Theory of CGMF}

In the SGMF Hamiltonian Eq.~(\ref{ham_ss}), we decouple the hopping terms 
between two neighbouring sites by considering the mean field or SF order 
parameter $\phi$. Thus, we could write the Hamiltonian of the entire 
system as the sum of Hamiltonians of individual sites and implement it as a
site wise computations.  However, this approximation is inadequate 
to incorporate the correlation effects arising from the NN hopping. To remedy
this short coming, which assumes great importance to describe strongly 
correlated states like QH states, previous works have relied on 
CGMF \cite{natu_16}. To derive the CGMF Hamiltonian, we consider the entire
lattice size as $K \times L$, which we divide into $W$ clusters ($C$) of 
size $M \times N$, i.e., $W = (K\times L)/(M\times N)$. The case of 
$M = N = 1$ corresponds to the SGMF theory. In CGMF Hamiltonian, the hopping 
term is decomposed into two parts. First is the actual hopping term in the 
internal link of the cluster($\delta$C) and the second term takes care of the 
boundary via mean fields. Our recent study \cite{bai_18} describes the 
decomposition of hopping term and CGMF method more clearly. After 
decomposition, the Hamiltonian for a single cluster($C$) is expressed in 
the following way 
\begin{eqnarray}
 \hat{H}_C  &=& -\sum_{p, q \in C}\left[\left({\rm e}^{i2\pi\alpha q}J_x 
              \hat{b}_{p + 1, q}^{\dagger}\hat{b}_{p, q} + {\rm H.c.}\right)
               \right. \nonumber\\
              &&+ \left. \left(J_y \hat{b}_{p, q - 1}^{\dagger}\hat{b}_{p, q}
              + {\rm H.c.}\right)\right]
              \nonumber\\
              &&-\sum_{p, q\in \delta C}
              \left[\left({\rm e}^{i2\pi\alpha q}J_x 
              \langle a_{p,q}^{*} \rangle \hat{b}_{p, q} + {\rm H.c.}\right)
              \right. \nonumber \\
              &&+ \left. \left(J_y \langle a_{p,q}^{*} \rangle\hat{b}_{p, q}
              + {\rm H.c.}\right)\right]
              \nonumber\\
              &&+\sum_{p, q \in C}
              \left[\frac{U}{2}\hat{n}_{p, q}(\hat{n}_{p, q}-1) - 
              \mu\hat{n}_{p, q}\right], 
\label{cg_hamil}         
\end{eqnarray}
where $\langle a_{p,q} \rangle  = \sum_{p^{'},q^{'}\in \not C} \langle 
b_{p^{'},q^{'}}\rangle$. Then, we use Gutzwiller ansatz and the local
cluster wavefunction in a Fock basis can be expressed as
\begin{equation}
\ket{\psi_c} = \sum_{n_1,n_2,\ldots,n_{MN}}
  C_{n_1,n_2..,n_{MN}}\ket{n_1,n_2,\ldots,n_{MN}},
\end{equation} 
with $n_i$ being the index of the occupation number state of $i$th lattice 
site within the cluster, and $C_{n_1,n_2,\ldots,n_{MN}}$ is the amplitude 
of the cluster Fock state $|n_1,n_2,\ldots, n_{MN}\rangle$. 
Here also, the total Hamiltonian of the system can be written as the sum of 
all the individual cluster Hamiltonians \cite{luhmann_13}. 
The SF order parameter $\phi$ is computed for each cluster in the similar way 
as discussed in SGMF method. The next step is to find the ground state and we 
adopt the similar process as is described in SGMF theory. 
We take the initial solution for $\phi$, construct the Hamiltonian matrix 
elements for a single cluster and diagonalize it. After diagonalization, 
we consider the lowest ground state $\ket{\psi_c}$ for the cluster and 
calculate the new $\phi$ and repeat the cycle until we get the converged 
solution. Here, it is worth to be mention that in case of CGMF, the 
convergence is very sensitive to the initial 
conditions \cite{natu_16,kuno_17} and we use the method of 
successive over-relaxation for the better convergence \cite{barrett_94}.

%
\begin{figure}[t]
  \includegraphics[width=8.3cm]{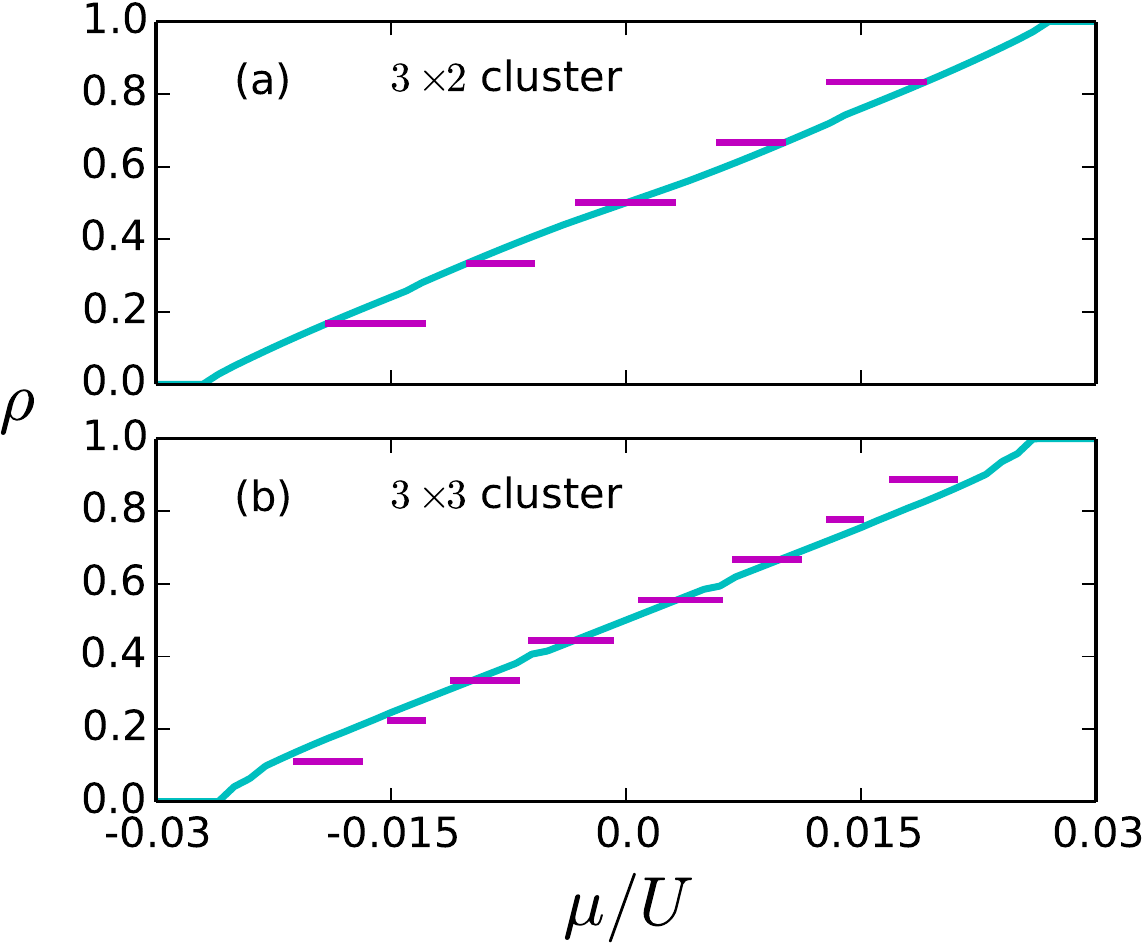}
  \caption{The variation in the number density $\rho$ for $\alpha = 1/3$
           as function of $\mu$. The states in SF phase are compressible and 
           have non-zero superfluid order parameter $\phi$. As a result, 
           $\rho$ varies linearly with $\mu$ and the green curve represents 
           the SF states. For specific values of filling factor $\nu$ there are 
           states with constant $\rho$, represented by the blue lines, and 
           these correspond to the QH states. 
           (a) Results from $3\times2$ cluster, the plateaus or 
           the constant $\rho$ values correspond to $\nu = n/2, n =1,2,..,5$ 
           and the corresponding $\rho$ values are $n/6$.
           (b) Results from $3\times3$, the platues correspond to 
           $\nu = n/3, n = 1,2,..,8$ and the corresponding $\rho$ values 
           are $n/9$.}
  \label{kappa_alpha1b3}
\end{figure}
\begin{figure}[t]
  \includegraphics[width=8.3cm]{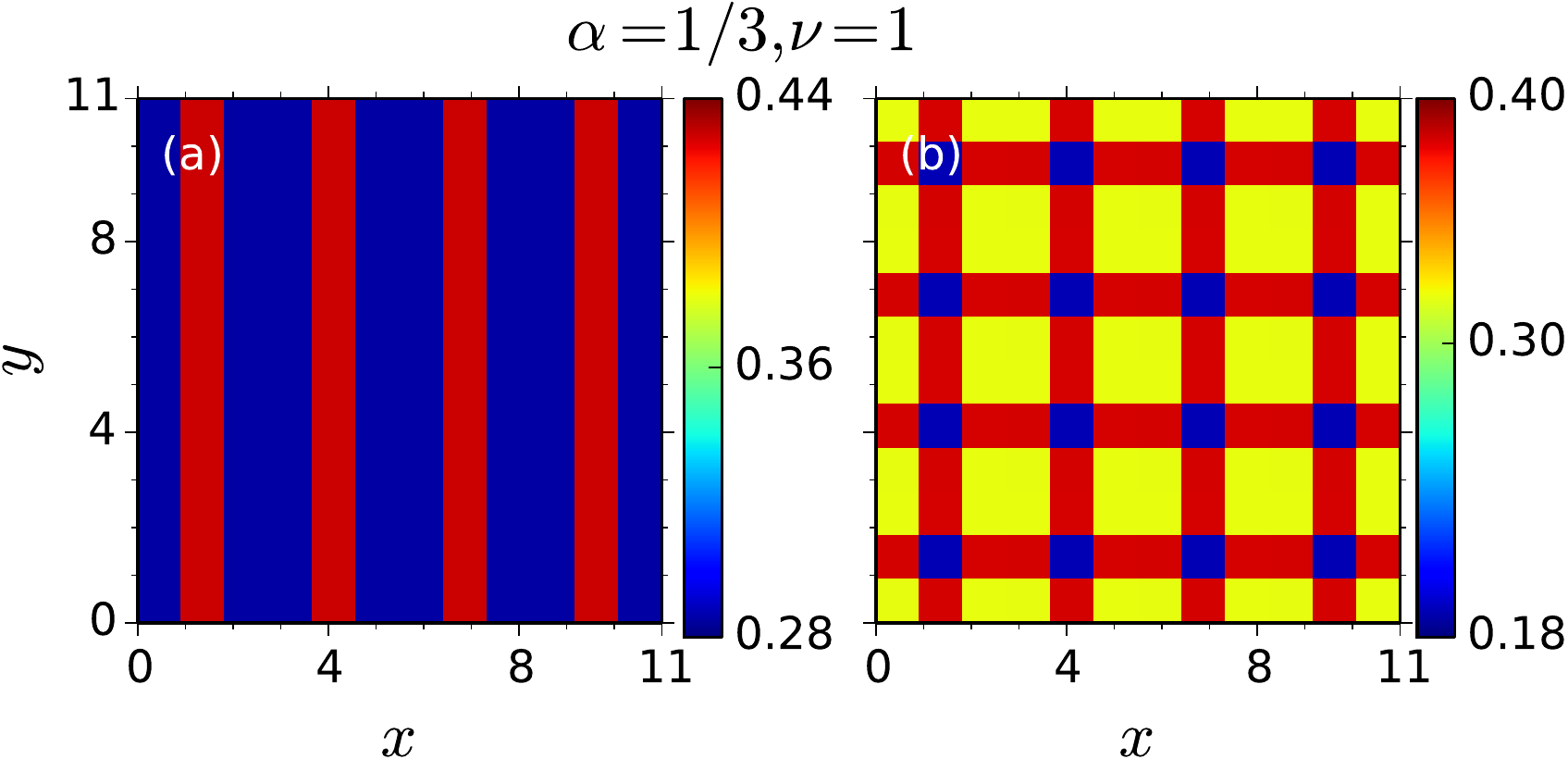}
  \caption{The IQH state for $\alpha = 1/3$ and $\nu = 1$ with (a) $3\times2$ 
           cluster and (b) with $3\times3$ cluster. The IQH state switches from
           stripe to checkerboard geometry with the mentioned cluster sizes.}
  \label{hs_nu_1_3b2_3b3}
\end{figure}
%


\section{Results and Discussions}

\begin{figure}[t]
  \includegraphics[width=8.3cm]{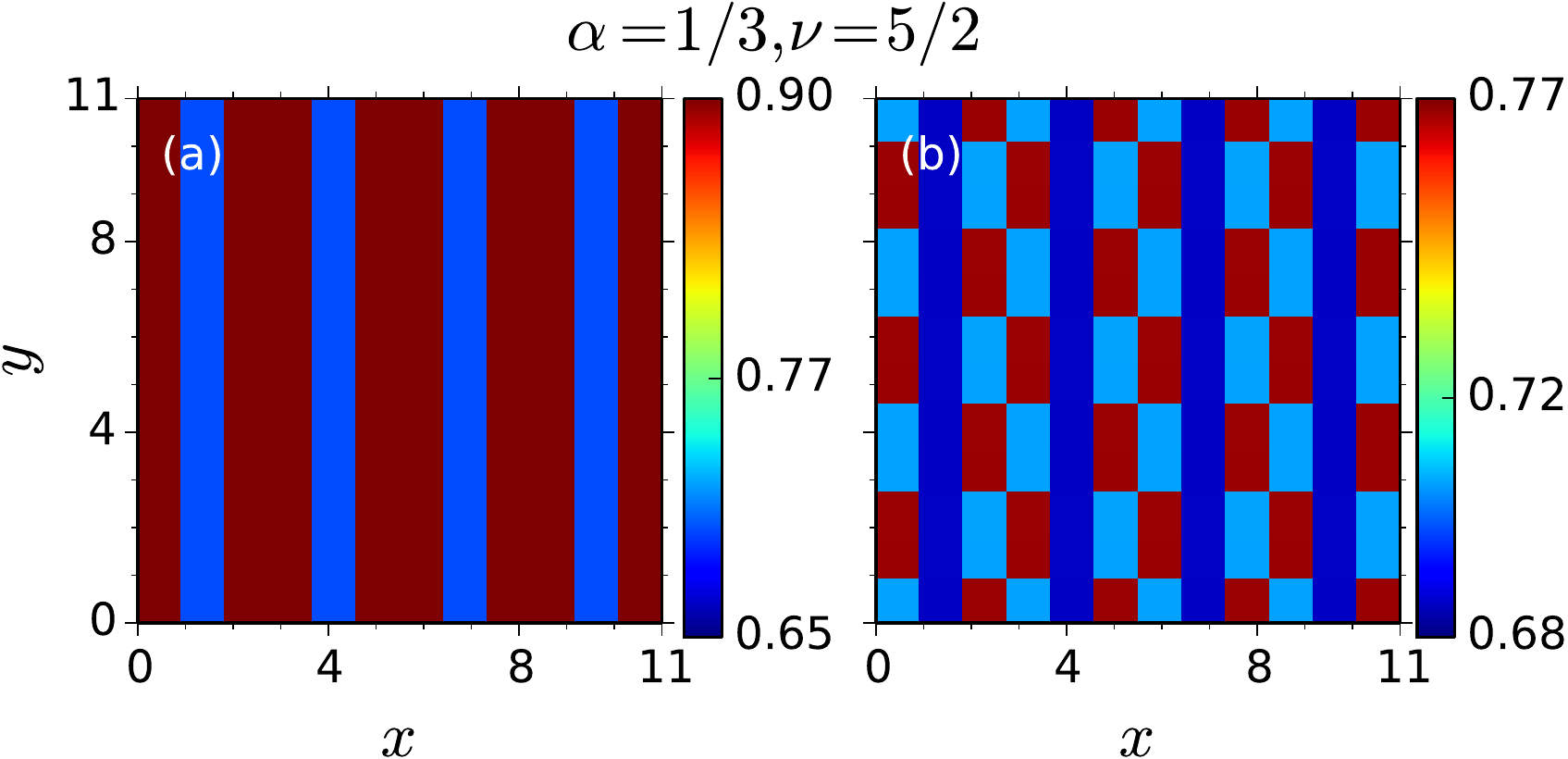}
  \caption{(a) The FQH state with $\alpha = 1/3$ 
           and $\nu = 5/2$ with $3\times2$ cluster, it has a stripe pattern
           in the density with vanishing SF order parameter. 
           (b) The analogous SF state with zigzag pattern in 
           the density as well as in the SF order parameter.}
  \label{nu_5b2_3b2}
\end{figure}

We start our computations by considering a single cluster for 
$\alpha = 1/3$, with $3\times2$ and $3\times3$ clusters. We choose
these clusters as Hamiltonian becomes periodic over a $3 \times 1$ 
magnetic unit cell in the Landau gauge for this flux value of $\alpha$. 
We obtain QH states from the CGMF and characterize them based on the 
compressibility $\kappa = \partial \rho /\partial \mu$, where the 
density for the cluster is 
$\rho = \sum_j\bra{\psi_c} \hat{n}_j \ket{\psi_c}/(K\times L)$.
As the QH states are incompressible $\kappa=0$ for these states, 
and $\kappa$ is finite for the compressible SF states. 
Therefore, $\rho(\mu)$ of QH states has plateaus for different fillings 
$\nu$ and $\rho(\mu)$ is linear for the SF phase. In the 
Fig.~\ref{kappa_alpha1b3}, the plateaus corresponding to constant $\rho$ 
indicate the existence for the QH states. Our computations, as mentioned
earlier, are in the hard-core boson limit where $\rho < 1$. We obtain the QH 
states at $\nu = n/2$, with $n =1,2,..,5$ by taking the $3\times2$ cluster, 
and at $\nu = n/3$, with $n = 1,2,..,8$ by taking the $3\times3$ cluster. The 
QH states are enhanced with the larger cluster size as mentioned above. 
Here, $3\times3$ cluster is close to exact diagonalization (ED) as the central 
lattice site has exact hopping contributions from the nearest neighbor sites. 
And, indeed, the diagonalization of the cluster can be transformed into ED with
minor modifications in the computations of the Hamiltonian matrix elements. One 
main reason for enhancement in the QH states with $3\times3$ cluster is that 
it describes correlations effects more accurately compared to the $3\times2$ 
cluster and hence, the results are more accurate. Further, we show the density 
plots for the QH and SF states for the larger lattice system. For this, we take 
$12 \times 12$ lattice sites and $J/U = 0.01$. This system size, 
and hopping energies are kept the same for all the QH and SF states discussed
in the rest of the manuscript. The IQH state at filling $\nu = 1$, 
$\mu/U = -0.008$ with density $\rho = 1/3$ is shown in 
Fig.~\ref{hs_nu_1_3b2_3b3}. The IQH state has stripe pattern with $3\times2$ 
cluster and transforms into checkerboard pattern with $3\times3$ cluster. The 
transformation from the stripe to checkerboard pattern is observed for the 
other IQH state of $\nu = 2$ as well. We observe that all the 
IQH and FQH states have stripe pattern except the $\nu = 3/2$ state, which 
has homogeneous density with $\rho = 0.5$ with $3\times2$ cluster. And the 
corresponding SF states have a zigzag pattern in the density $\rho$ and 
in the SF order parameter $\phi$. One of the FQH state for $\nu = 5/2$ and 
corresponding SF state is shown in the Fig.~\ref{nu_5b2_3b2}. 
%
\begin{figure}[t]
  \includegraphics[width=8.3cm]{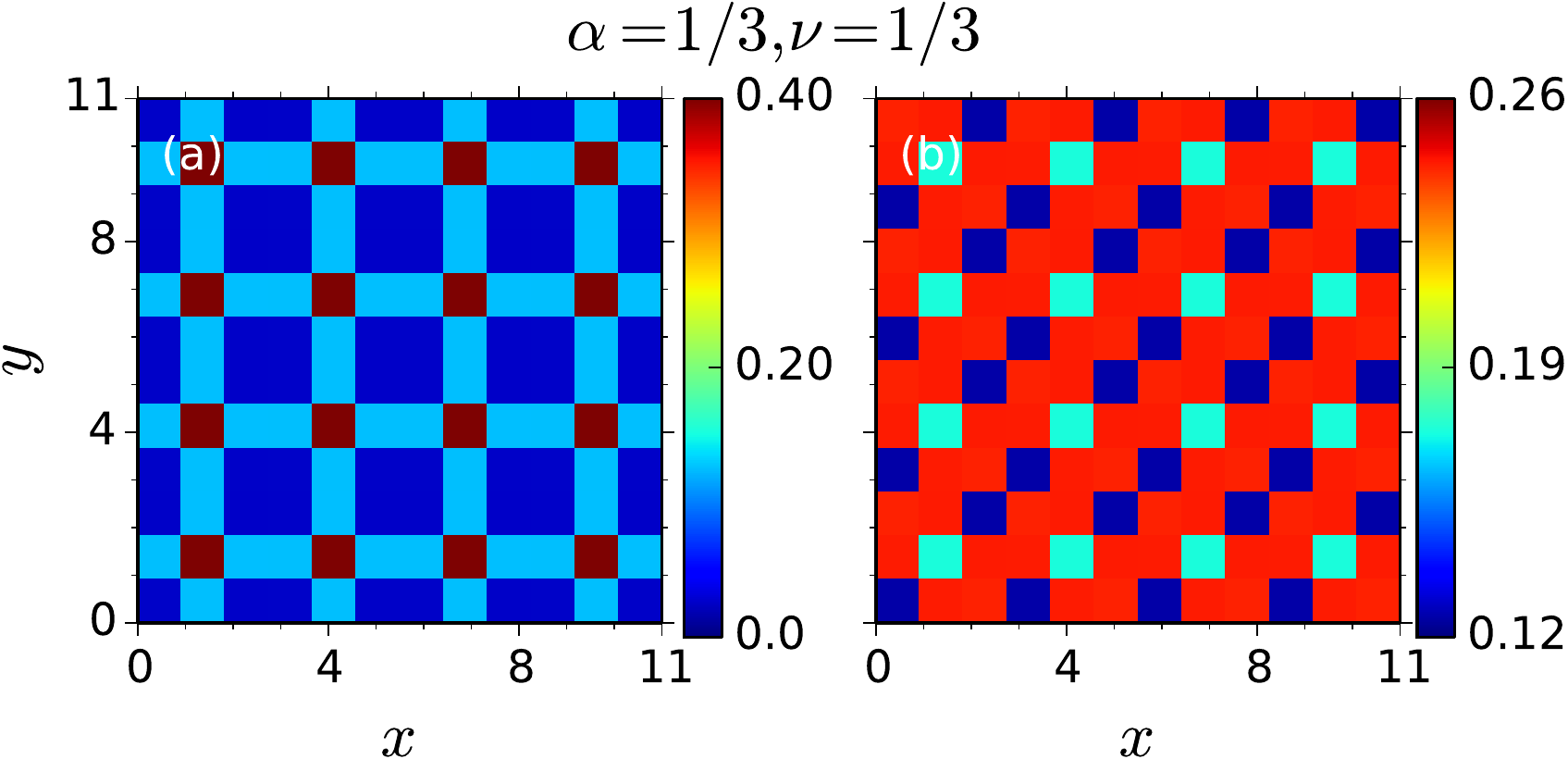}
  \caption{(a) The represents FQH state with $\alpha = 1/3$ 
           and $\nu = 1/3$ with $3\times3$ cluster with a checkerboard pattern
           in the density and vanishing SF order parameter. 
           (b) The analogous SF state with diagonal stripe pattern 
           in the density as well as in the SF order parameter.}
  \label{nu_1b3_3b3}
\end{figure}
%
As discussed earlier, we do not observe the half integer FQH states with
$3\times3$ cluster, but do observe the FQH states at the one third fillings.
One of the FQH and SF state with $3\times3$ cluster for $\nu = 1/3$ is 
shown in the Fig.~\ref{nu_1b3_3b3}. Here, with $3\times3$ cluster, we observe 
all the FQH states have checkerboard pattern and all the SF states have the
diagonal stripe pattern. We find that in all the cases SF states is the 
ground state and QH state is metastable state.


\section{Conclusion}

We obtain QH states by considering the two cluster sizes as $3 \times 2$
and $3 \times 3$ in the CGMF theory. With the larger cluster $3 \times 3$, we 
approach ED as the CGMF provides exact description of the hopping term for
the central lattice site. We obtain QH states with fillings 
$\nu = n/2$, $n = 1,2,..,5$ and corresponding density $\rho = n/6$ with 
$3\times 2$ cluster size. However, with $3\times 3$ cluster, we obtain
a larger set of QH states with fillings $\nu = n/3$, $n = 1,2,..,8$
and corresponding density $\rho = n/9$. We also observe the competing SF 
states corresponding to all the QH states. We find that the SF state is
the ground state and QH state is metastable state in all the cases.
We have demonstrated that the QH states change geometry from stripe to 
checkerboard by switching $3\times 2$ to $3\times 3$ cluster size. On the 
other hand, SF state has zigzag pattern with both the cluster sizes, specially 
with $3\times3$ the zigzag pattern is equivalent to diagonal stripe pattern. 
Thus, we have established that to obtain correct density pattern of the 
QH states, it is essential to consider larger cluster sizes in the CGMF
theory.


\begin{acknowledgments}
The results presented in the paper are based on the computations
using Vikram-100, the 100TFLOP HPC Cluster at Physical Research Laboratory, 
Ahmedabad, India. We thank Arko Roy, S. Gautam and S. A. Silotri for valuable 
discussions.
\end{acknowledgments}
\bibliography{ref}{}

\begin{thebibliography}{44}%
\makeatletter
\providecommand \@ifxundefined [1]{%
 \@ifx{#1\undefined}
}%
\providecommand \@ifnum [1]{%
 \ifnum #1\expandafter \@firstoftwo
 \else \expandafter \@secondoftwo
 \fi
}%
\providecommand \@ifx [1]{%
 \ifx #1\expandafter \@firstoftwo
 \else \expandafter \@secondoftwo
 \fi
}%
\providecommand \natexlab [1]{#1}%
\providecommand \enquote  [1]{``#1''}%
\providecommand \bibnamefont  [1]{#1}%
\providecommand \bibfnamefont [1]{#1}%
\providecommand \citenamefont [1]{#1}%
\providecommand \href@noop [0]{\@secondoftwo}%
\providecommand \href [0]{\begingroup \@sanitize@url \@href}%
\providecommand \@href[1]{\@@startlink{#1}\@@href}%
\providecommand \@@href[1]{\endgroup#1\@@endlink}%
\providecommand \@sanitize@url [0]{\catcode `\\12\catcode `\$12\catcode
  `\&12\catcode `\#12\catcode `\^12\catcode `\_12\catcode `\%12\relax}%
\providecommand \@@startlink[1]{}%
\providecommand \@@endlink[0]{}%
\providecommand \url  [0]{\begingroup\@sanitize@url \@url }%
\providecommand \@url [1]{\endgroup\@href {#1}{\urlprefix }}%
\providecommand \urlprefix  [0]{URL }%
\providecommand \Eprint [0]{\href }%
\providecommand \doibase [0]{http://dx.doi.org/}%
\providecommand \selectlanguage [0]{\@gobble}%
\providecommand \bibinfo  [0]{\@secondoftwo}%
\providecommand \bibfield  [0]{\@secondoftwo}%
\providecommand \translation [1]{[#1]}%
\providecommand \BibitemOpen [0]{}%
\providecommand \bibitemStop [0]{}%
\providecommand \bibitemNoStop [0]{.\EOS\space}%
\providecommand \EOS [0]{\spacefactor3000\relax}%
\providecommand \BibitemShut  [1]{\csname bibitem#1\endcsname}%
\let\auto@bib@innerbib\@empty
\bibitem [{\citenamefont {Bloch}\ \emph {et~al.}(2008)\citenamefont {Bloch},
  \citenamefont {Dalibard},\ and\ \citenamefont {Zwerger}}]{bloch_08}%
  \BibitemOpen
  \bibfield  {author} {\bibinfo {author} {\bibfnamefont {I.}~\bibnamefont
  {Bloch}}, \bibinfo {author} {\bibfnamefont {J.}~\bibnamefont {Dalibard}}, \
  and\ \bibinfo {author} {\bibfnamefont {W.}~\bibnamefont {Zwerger}},\
  }\bibfield  {title} {\enquote {\bibinfo {title} {Many-body physics with
  ultracold gases},}\ }\href {\doibase 10.1103/RevModPhys.80.885} {\bibfield
  {journal} {\bibinfo  {journal} {Rev. Mod. Phys.}\ }\textbf {\bibinfo {volume}
  {80}},\ \bibinfo {pages} {885} (\bibinfo {year} {2008})}\BibitemShut
  {NoStop}%
\bibitem [{\citenamefont {Anderson}\ and\ \citenamefont
  {Kasevich}(1998)}]{anderson_98}%
  \BibitemOpen
  \bibfield  {author} {\bibinfo {author} {\bibfnamefont {B.~P.}\ \bibnamefont
  {Anderson}}\ and\ \bibinfo {author} {\bibfnamefont {M.~A.}\ \bibnamefont
  {Kasevich}},\ }\bibfield  {title} {\enquote {\bibinfo {title} {Macroscopic
  quantum interference from atomic tunnel arrays},}\ }\href {\doibase
  10.1126/science.282.5394.1686} {\bibfield  {journal} {\bibinfo  {journal}
  {Science}\ }\textbf {\bibinfo {volume} {282}},\ \bibinfo {pages} {1686}
  (\bibinfo {year} {1998})}\BibitemShut {NoStop}%
\bibitem [{\citenamefont {Greiner}\ \emph {et~al.}(2001)\citenamefont
  {Greiner}, \citenamefont {Bloch}, \citenamefont {Mandel}, \citenamefont
  {H\"ansch},\ and\ \citenamefont {Esslinger}}]{greiner_01}%
  \BibitemOpen
  \bibfield  {author} {\bibinfo {author} {\bibfnamefont {M.}~\bibnamefont
  {Greiner}}, \bibinfo {author} {\bibfnamefont {I.}~\bibnamefont {Bloch}},
  \bibinfo {author} {\bibfnamefont {O.}~\bibnamefont {Mandel}}, \bibinfo
  {author} {\bibfnamefont {T.~W.}\ \bibnamefont {H\"ansch}}, \ and\ \bibinfo
  {author} {\bibfnamefont {T.}~\bibnamefont {Esslinger}},\ }\bibfield  {title}
  {\enquote {\bibinfo {title} {Exploring phase coherence in a 2{D} lattice of
  {B}ose-{E}instein condensates},}\ }\href {\doibase
  10.1103/PhysRevLett.87.160405} {\bibfield  {journal} {\bibinfo  {journal}
  {Phys. Rev. Lett.}\ }\textbf {\bibinfo {volume} {87}},\ \bibinfo {pages}
  {160405} (\bibinfo {year} {2001})}\BibitemShut {NoStop}%
\bibitem [{\citenamefont {Greiner}\ \emph {et~al.}(2002)\citenamefont
  {Greiner}, \citenamefont {Mandel}, \citenamefont {Esslinger}, \citenamefont
  {H\"ansch},\ and\ \citenamefont {Bloch}}]{greiner_02}%
  \BibitemOpen
  \bibfield  {author} {\bibinfo {author} {\bibfnamefont {M.}~\bibnamefont
  {Greiner}}, \bibinfo {author} {\bibfnamefont {O.}~\bibnamefont {Mandel}},
  \bibinfo {author} {\bibfnamefont {T.}~\bibnamefont {Esslinger}}, \bibinfo
  {author} {\bibfnamefont {T.~W.}\ \bibnamefont {H\"ansch}}, \ and\ \bibinfo
  {author} {\bibfnamefont {I.}~\bibnamefont {Bloch}},\ }\bibfield  {title}
  {\enquote {\bibinfo {title} {Quantum phase transition from a superfluid to a
  {M}ott insulator in a gas of ultracold atoms},}\ }\href {\doibase
  10.1038/415039a} {\bibfield  {journal} {\bibinfo  {journal} {Nature
  (London)}\ }\textbf {\bibinfo {volume} {415}},\ \bibinfo {pages} {39}
  (\bibinfo {year} {2002})}\BibitemShut {NoStop}%
\bibitem [{\citenamefont {Lewenstein}\ \emph {et~al.}(2007)\citenamefont
  {Lewenstein}, \citenamefont {Sanpera}, \citenamefont {Ahufinger},
  \citenamefont {Damski}, \citenamefont {Sen(De)},\ and\ \citenamefont
  {Sen}}]{lewenstein_07}%
  \BibitemOpen
  \bibfield  {author} {\bibinfo {author} {\bibfnamefont {M.}~\bibnamefont
  {Lewenstein}}, \bibinfo {author} {\bibfnamefont {A.}~\bibnamefont {Sanpera}},
  \bibinfo {author} {\bibfnamefont {V.}~\bibnamefont {Ahufinger}}, \bibinfo
  {author} {\bibfnamefont {B.}~\bibnamefont {Damski}}, \bibinfo {author}
  {\bibfnamefont {A.}~\bibnamefont {Sen(De)}}, \ and\ \bibinfo {author}
  {\bibfnamefont {U.}~\bibnamefont {Sen}},\ }\bibfield  {title} {\enquote
  {\bibinfo {title} {Ultracold atomic gases in optical lattices: mimicking
  condensed matter physics and beyond},}\ }\href {\doibase
  10.1080/00018730701223200} {\bibfield  {journal} {\bibinfo  {journal} {Adv.
  Phys.}\ }\textbf {\bibinfo {volume} {56}},\ \bibinfo {pages} {243} (\bibinfo
  {year} {2007})}\BibitemShut {NoStop}%
\bibitem [{\citenamefont {Aidelsburger}\ \emph {et~al.}(2013)\citenamefont
  {Aidelsburger}, \citenamefont {Atala}, \citenamefont {Lohse}, \citenamefont
  {Barreiro}, \citenamefont {Paredes},\ and\ \citenamefont
  {Bloch}}]{aidelsburger_13}%
  \BibitemOpen
  \bibfield  {author} {\bibinfo {author} {\bibfnamefont {M.}~\bibnamefont
  {Aidelsburger}}, \bibinfo {author} {\bibfnamefont {M.}~\bibnamefont {Atala}},
  \bibinfo {author} {\bibfnamefont {M.}~\bibnamefont {Lohse}}, \bibinfo
  {author} {\bibfnamefont {J.~T.}\ \bibnamefont {Barreiro}}, \bibinfo {author}
  {\bibfnamefont {B.}~\bibnamefont {Paredes}}, \ and\ \bibinfo {author}
  {\bibfnamefont {I.}~\bibnamefont {Bloch}},\ }\bibfield  {title} {\enquote
  {\bibinfo {title} {Realization of the {H}ofstadter {H}amiltonian with
  ultracold atoms in optical lattices},}\ }\href {\doibase
  10.1103/PhysRevLett.111.185301} {\bibfield  {journal} {\bibinfo  {journal}
  {Phys. Rev. Lett.}\ }\textbf {\bibinfo {volume} {111}},\ \bibinfo {pages}
  {185301} (\bibinfo {year} {2013})}\BibitemShut {NoStop}%
\bibitem [{\citenamefont {Miyake}\ \emph {et~al.}(2013)\citenamefont {Miyake},
  \citenamefont {Siviloglou}, \citenamefont {Kennedy}, \citenamefont {Burton},\
  and\ \citenamefont {Ketterle}}]{miyake_13}%
  \BibitemOpen
  \bibfield  {author} {\bibinfo {author} {\bibfnamefont {H.}~\bibnamefont
  {Miyake}}, \bibinfo {author} {\bibfnamefont {G.~A.}\ \bibnamefont
  {Siviloglou}}, \bibinfo {author} {\bibfnamefont {C.~J.}\ \bibnamefont
  {Kennedy}}, \bibinfo {author} {\bibfnamefont {W.~C.}\ \bibnamefont {Burton}},
  \ and\ \bibinfo {author} {\bibfnamefont {W.}~\bibnamefont {Ketterle}},\
  }\bibfield  {title} {\enquote {\bibinfo {title} {Realizing the {H}arper
  {H}amiltonian with laser-assisted tunneling in optical lattices},}\ }\href
  {\doibase 10.1103/PhysRevLett.111.185302} {\bibfield  {journal} {\bibinfo
  {journal} {Phys. Rev. Lett.}\ }\textbf {\bibinfo {volume} {111}},\ \bibinfo
  {pages} {185302} (\bibinfo {year} {2013})}\BibitemShut {NoStop}%
\bibitem [{\citenamefont {Harper}(1955)}]{harper_55}%
  \BibitemOpen
  \bibfield  {author} {\bibinfo {author} {\bibfnamefont {P.~G.}\ \bibnamefont
  {Harper}},\ }\bibfield  {title} {\enquote {\bibinfo {title} {Single band
  motion of conduction electrons in a uniform magnetic field},}\ }\href
  {http://iopscience.iop.org/0370-1298/68/10/304} {\bibfield  {journal}
  {\bibinfo  {journal} {Proc. Phys. Soc. A}\ }\textbf {\bibinfo {volume}
  {68}},\ \bibinfo {pages} {874} (\bibinfo {year} {1955})}\BibitemShut
  {NoStop}%
\bibitem [{\citenamefont {Hofstadter}(1976)}]{hofstadter_76}%
  \BibitemOpen
  \bibfield  {author} {\bibinfo {author} {\bibfnamefont {D.~R.}\ \bibnamefont
  {Hofstadter}},\ }\bibfield  {title} {\enquote {\bibinfo {title} {Energy
  levels and wave functions of {B}loch electrons in rational and irrational
  magnetic fields},}\ }\href {\doibase 10.1103/PhysRevB.14.2239} {\bibfield
  {journal} {\bibinfo  {journal} {Phys. Rev. B}\ }\textbf {\bibinfo {volume}
  {14}},\ \bibinfo {pages} {2239} (\bibinfo {year} {1976})}\BibitemShut
  {NoStop}%
\bibitem [{\citenamefont {Dean}\ \emph {et~al.}(2013)\citenamefont {Dean},
  \citenamefont {Wang}, \citenamefont {Maher}, \citenamefont {Forsythe},
  \citenamefont {Ghahari}, \citenamefont {Gao}, \citenamefont {Katoch},
  \citenamefont {Ishigami}, \citenamefont {Moon}, \citenamefont {Koshino},
  \citenamefont {Taniguchi}, \citenamefont {Watanabe}, \citenamefont {Shepard},
  \citenamefont {Hone},\ and\ \citenamefont {Kim}}]{dean_13}%
  \BibitemOpen
  \bibfield  {author} {\bibinfo {author} {\bibfnamefont {C.~R.}\ \bibnamefont
  {Dean}}, \bibinfo {author} {\bibfnamefont {L.}~\bibnamefont {Wang}}, \bibinfo
  {author} {\bibfnamefont {P.}~\bibnamefont {Maher}}, \bibinfo {author}
  {\bibfnamefont {C.}~\bibnamefont {Forsythe}}, \bibinfo {author}
  {\bibfnamefont {F.}~\bibnamefont {Ghahari}}, \bibinfo {author} {\bibfnamefont
  {Y.}~\bibnamefont {Gao}}, \bibinfo {author} {\bibfnamefont {J.}~\bibnamefont
  {Katoch}}, \bibinfo {author} {\bibfnamefont {M.}~\bibnamefont {Ishigami}},
  \bibinfo {author} {\bibfnamefont {P.}~\bibnamefont {Moon}}, \bibinfo {author}
  {\bibfnamefont {M.}~\bibnamefont {Koshino}}, \bibinfo {author} {\bibfnamefont
  {T.}~\bibnamefont {Taniguchi}}, \bibinfo {author} {\bibfnamefont
  {K.}~\bibnamefont {Watanabe}}, \bibinfo {author} {\bibfnamefont {K.~L.}\
  \bibnamefont {Shepard}}, \bibinfo {author} {\bibfnamefont {J.}~\bibnamefont
  {Hone}}, \ and\ \bibinfo {author} {\bibfnamefont {P.}~\bibnamefont {Kim}},\
  }\bibfield  {title} {\enquote {\bibinfo {title} {Hofstadter’s butterfly and
  the fractal quantum hall effect in moiré superlattices},}\ }\href {\doibase
  10.1038/nature12186} {\bibfield  {journal} {\bibinfo  {journal} {Nature}\
  }\textbf {\bibinfo {volume} {497}},\ \bibinfo {pages} {598} (\bibinfo {year}
  {2013})}\BibitemShut {NoStop}%
\bibitem [{\citenamefont {Jaksch}\ and\ \citenamefont
  {Zoller}(2003)}]{jaksch_03}%
  \BibitemOpen
  \bibfield  {author} {\bibinfo {author} {\bibfnamefont {D.}~\bibnamefont
  {Jaksch}}\ and\ \bibinfo {author} {\bibfnamefont {P.}~\bibnamefont
  {Zoller}},\ }\bibfield  {title} {\enquote {\bibinfo {title} {Creation of
  effective magnetic fields in optical lattices: the {H}ofstadter butterfly for
  cold neutral atoms},}\ }\href {\doibase
  https://doi.org/10.1088/1367-2630/5/1/356} {\bibfield  {journal} {\bibinfo
  {journal} {New J. Phys.}\ }\textbf {\bibinfo {volume} {5}},\ \bibinfo {pages}
  {56} (\bibinfo {year} {2003})}\BibitemShut {NoStop}%
\bibitem [{\citenamefont {Dalibard}\ \emph {et~al.}(2011)\citenamefont
  {Dalibard}, \citenamefont {Gerbier}, \citenamefont
  {Juzeli\ifmmode~\bar{u}\else \={u}\fi{}nas},\ and\ \citenamefont
  {\"Ohberg}}]{dalibard_11}%
  \BibitemOpen
  \bibfield  {author} {\bibinfo {author} {\bibfnamefont {J.}~\bibnamefont
  {Dalibard}}, \bibinfo {author} {\bibfnamefont {F.}~\bibnamefont {Gerbier}},
  \bibinfo {author} {\bibfnamefont {G.}~\bibnamefont
  {Juzeli\ifmmode~\bar{u}\else \={u}\fi{}nas}}, \ and\ \bibinfo {author}
  {\bibfnamefont {P.}~\bibnamefont {\"Ohberg}},\ }\bibfield  {title} {\enquote
  {\bibinfo {title} {Colloquium: Artificial gauge potentials for neutral
  atoms},}\ }\href {\doibase 10.1103/RevModPhys.83.1523} {\bibfield  {journal}
  {\bibinfo  {journal} {Rev. Mod. Phys.}\ }\textbf {\bibinfo {volume} {83}},\
  \bibinfo {pages} {1523} (\bibinfo {year} {2011})}\BibitemShut {NoStop}%
\bibitem [{\citenamefont {Lin}\ \emph {et~al.}(2011)\citenamefont {Lin},
  \citenamefont {Compton}, \citenamefont {Jimenez-Garcia}, \citenamefont
  {Phillips}, \citenamefont {Porto},\ and\ \citenamefont {Spielman}}]{lin_11}%
  \BibitemOpen
  \bibfield  {author} {\bibinfo {author} {\bibfnamefont {Y-J.}\ \bibnamefont
  {Lin}}, \bibinfo {author} {\bibfnamefont {R.~L.}\ \bibnamefont {Compton}},
  \bibinfo {author} {\bibfnamefont {K.}~\bibnamefont {Jimenez-Garcia}},
  \bibinfo {author} {\bibfnamefont {W.~D.}\ \bibnamefont {Phillips}}, \bibinfo
  {author} {\bibfnamefont {J.~V.}\ \bibnamefont {Porto}}, \ and\ \bibinfo
  {author} {\bibfnamefont {I.~B.}\ \bibnamefont {Spielman}},\ }\bibfield
  {title} {\enquote {\bibinfo {title} {A synthetic electric force acting on
  neutral atoms},}\ }\href {\doibase 10.1038/nphys1954} {\bibfield  {journal}
  {\bibinfo  {journal} {Nat. Phys.}\ }\textbf {\bibinfo {volume} {7}},\
  \bibinfo {pages} {531} (\bibinfo {year} {2011})}\BibitemShut {NoStop}%
\bibitem [{\citenamefont {Lin}\ \emph {et~al.}(2009)\citenamefont {Lin},
  \citenamefont {Compton}, \citenamefont {Perry}, \citenamefont {Phillips},
  \citenamefont {Porto},\ and\ \citenamefont {Spielman}}]{lin_09a}%
  \BibitemOpen
  \bibfield  {author} {\bibinfo {author} {\bibfnamefont {Y.-J.}\ \bibnamefont
  {Lin}}, \bibinfo {author} {\bibfnamefont {R.~L.}\ \bibnamefont {Compton}},
  \bibinfo {author} {\bibfnamefont {A.~R.}\ \bibnamefont {Perry}}, \bibinfo
  {author} {\bibfnamefont {W.~D.}\ \bibnamefont {Phillips}}, \bibinfo {author}
  {\bibfnamefont {J.~V.}\ \bibnamefont {Porto}}, \ and\ \bibinfo {author}
  {\bibfnamefont {I.~B.}\ \bibnamefont {Spielman}},\ }\bibfield  {title}
  {\enquote {\bibinfo {title} {{B}ose-{E}instein condensate in a uniform
  light-induced vector potential},}\ }\href {\doibase
  10.1103/PhysRevLett.102.130401} {\bibfield  {journal} {\bibinfo  {journal}
  {Phys. Rev. Lett.}\ }\textbf {\bibinfo {volume} {102}},\ \bibinfo {pages}
  {130401} (\bibinfo {year} {2009})}\BibitemShut {NoStop}%
\bibitem [{\citenamefont {Jim\'enez-Garc\'{\i}a}\ \emph
  {et~al.}(2012)\citenamefont {Jim\'enez-Garc\'{\i}a}, \citenamefont {LeBlanc},
  \citenamefont {Williams}, \citenamefont {Beeler}, \citenamefont {Perry},\
  and\ \citenamefont {Spielman}}]{garcia_12}%
  \BibitemOpen
  \bibfield  {author} {\bibinfo {author} {\bibfnamefont {K.}~\bibnamefont
  {Jim\'enez-Garc\'{\i}a}}, \bibinfo {author} {\bibfnamefont {L.~J.}\
  \bibnamefont {LeBlanc}}, \bibinfo {author} {\bibfnamefont {R.~A.}\
  \bibnamefont {Williams}}, \bibinfo {author} {\bibfnamefont {M.~C.}\
  \bibnamefont {Beeler}}, \bibinfo {author} {\bibfnamefont {A.~R.}\
  \bibnamefont {Perry}}, \ and\ \bibinfo {author} {\bibfnamefont {I.~B.}\
  \bibnamefont {Spielman}},\ }\bibfield  {title} {\enquote {\bibinfo {title}
  {Peierls substitution in an engineered lattice potential},}\ }\href {\doibase
  10.1103/PhysRevLett.108.225303} {\bibfield  {journal} {\bibinfo  {journal}
  {Phys. Rev. Lett.}\ }\textbf {\bibinfo {volume} {108}},\ \bibinfo {pages}
  {225303} (\bibinfo {year} {2012})}\BibitemShut {NoStop}%
\bibitem [{\citenamefont {Peierls}(1933)}]{peierls_33}%
  \BibitemOpen
  \bibfield  {author} {\bibinfo {author} {\bibfnamefont {R.~E.}\ \bibnamefont
  {Peierls}},\ }\bibfield  {title} {\enquote {\bibinfo {title} {On the theory
  of diamagnetism of conduction electrons},}\ }\href {\doibase
  10.1007/BF01342591} {\bibfield  {journal} {\bibinfo  {journal} {Z. Phys.}\
  }\textbf {\bibinfo {volume} {80}},\ \bibinfo {pages} {763} (\bibinfo {year}
  {1933})}\BibitemShut {NoStop}%
\bibitem [{\citenamefont {Palmer}\ and\ \citenamefont
  {Jaksch}(2006)}]{palmer_06}%
  \BibitemOpen
  \bibfield  {author} {\bibinfo {author} {\bibfnamefont {R.~N.}\ \bibnamefont
  {Palmer}}\ and\ \bibinfo {author} {\bibfnamefont {D.}~\bibnamefont
  {Jaksch}},\ }\bibfield  {title} {\enquote {\bibinfo {title} {High-field
  fractional quantum {H}all effect in optical lattices},}\ }\href {\doibase
  10.1103/PhysRevLett.96.180407} {\bibfield  {journal} {\bibinfo  {journal}
  {Phys. Rev. Lett.}\ }\textbf {\bibinfo {volume} {96}},\ \bibinfo {pages}
  {180407} (\bibinfo {year} {2006})}\BibitemShut {NoStop}%
\bibitem [{\citenamefont {S\o{}rensen}\ \emph {et~al.}(2005)\citenamefont
  {S\o{}rensen}, \citenamefont {Demler},\ and\ \citenamefont
  {Lukin}}]{sorensen_05}%
  \BibitemOpen
  \bibfield  {author} {\bibinfo {author} {\bibfnamefont {A.~S.}\ \bibnamefont
  {S\o{}rensen}}, \bibinfo {author} {\bibfnamefont {E.}~\bibnamefont {Demler}},
  \ and\ \bibinfo {author} {\bibfnamefont {M.~D.}\ \bibnamefont {Lukin}},\
  }\bibfield  {title} {\enquote {\bibinfo {title} {Fractional quantum {H}all
  states of atoms in optical lattices},}\ }\href {\doibase
  10.1103/PhysRevLett.94.086803} {\bibfield  {journal} {\bibinfo  {journal}
  {Phys. Rev. Lett.}\ }\textbf {\bibinfo {volume} {94}},\ \bibinfo {pages}
  {086803} (\bibinfo {year} {2005})}\BibitemShut {NoStop}%
\bibitem [{\citenamefont {Fisher}\ \emph {et~al.}(1989)\citenamefont {Fisher},
  \citenamefont {Weichman}, \citenamefont {Grinstein},\ and\ \citenamefont
  {Fisher}}]{fisher_89}%
  \BibitemOpen
  \bibfield  {author} {\bibinfo {author} {\bibfnamefont {M.~P.~A.}\
  \bibnamefont {Fisher}}, \bibinfo {author} {\bibfnamefont {P.~B.}\
  \bibnamefont {Weichman}}, \bibinfo {author} {\bibfnamefont {G.}~\bibnamefont
  {Grinstein}}, \ and\ \bibinfo {author} {\bibfnamefont {D.~S.}\ \bibnamefont
  {Fisher}},\ }\bibfield  {title} {\enquote {\bibinfo {title} {Boson
  localization and the superfluid-insulator transition},}\ }\href {\doibase
  10.1103/PhysRevB.40.546} {\bibfield  {journal} {\bibinfo  {journal} {Phys.
  Rev. B}\ }\textbf {\bibinfo {volume} {40}},\ \bibinfo {pages} {546} (\bibinfo
  {year} {1989})}\BibitemShut {NoStop}%
\bibitem [{\citenamefont {Jaksch}\ \emph {et~al.}(1998)\citenamefont {Jaksch},
  \citenamefont {Bruder}, \citenamefont {Cirac}, \citenamefont {Gardiner},\
  and\ \citenamefont {Zoller}}]{jaksch_98}%
  \BibitemOpen
  \bibfield  {author} {\bibinfo {author} {\bibfnamefont {D.}~\bibnamefont
  {Jaksch}}, \bibinfo {author} {\bibfnamefont {C.}~\bibnamefont {Bruder}},
  \bibinfo {author} {\bibfnamefont {J.~I.}\ \bibnamefont {Cirac}}, \bibinfo
  {author} {\bibfnamefont {C.~W.}\ \bibnamefont {Gardiner}}, \ and\ \bibinfo
  {author} {\bibfnamefont {P.}~\bibnamefont {Zoller}},\ }\bibfield  {title}
  {\enquote {\bibinfo {title} {Cold bosonic atoms in optical lattices},}\
  }\href {\doibase 10.1103/PhysRevLett.81.3108} {\bibfield  {journal} {\bibinfo
   {journal} {Phys. Rev. Lett.}\ }\textbf {\bibinfo {volume} {81}},\ \bibinfo
  {pages} {3108} (\bibinfo {year} {1998})}\BibitemShut {NoStop}%
\bibitem [{\citenamefont {St\"oferle}\ \emph {et~al.}(2004)\citenamefont
  {St\"oferle}, \citenamefont {Moritz}, \citenamefont {Schori}, \citenamefont
  {K\"ohl},\ and\ \citenamefont {Esslinger}}]{stoferle_04}%
  \BibitemOpen
  \bibfield  {author} {\bibinfo {author} {\bibfnamefont {T.}~\bibnamefont
  {St\"oferle}}, \bibinfo {author} {\bibfnamefont {H.}~\bibnamefont {Moritz}},
  \bibinfo {author} {\bibfnamefont {C.}~\bibnamefont {Schori}}, \bibinfo
  {author} {\bibfnamefont {M.}~\bibnamefont {K\"ohl}}, \ and\ \bibinfo {author}
  {\bibfnamefont {T.}~\bibnamefont {Esslinger}},\ }\bibfield  {title} {\enquote
  {\bibinfo {title} {Transition from a strongly interacting {1D} superfluid to
  a {M}ott insulator},}\ }\href {\doibase 10.1103/PhysRevLett.92.130403}
  {\bibfield  {journal} {\bibinfo  {journal} {Phys. Rev. Lett.}\ }\textbf
  {\bibinfo {volume} {92}},\ \bibinfo {pages} {130403} (\bibinfo {year}
  {2004})}\BibitemShut {NoStop}%
\bibitem [{\citenamefont {Freericks}\ and\ \citenamefont
  {Monien}(1996)}]{freericks_96}%
  \BibitemOpen
  \bibfield  {author} {\bibinfo {author} {\bibfnamefont {J.~K.}\ \bibnamefont
  {Freericks}}\ and\ \bibinfo {author} {\bibfnamefont {H.}~\bibnamefont
  {Monien}},\ }\bibfield  {title} {\enquote {\bibinfo {title} {Strong-coupling
  expansions for the pure and disordered bose- hubbard model},}\ }\href
  {\doibase 10.1103/PhysRevB.53.2691} {\bibfield  {journal} {\bibinfo
  {journal} {Phys. Rev. B}\ }\textbf {\bibinfo {volume} {53}},\ \bibinfo
  {pages} {2691--2700} (\bibinfo {year} {1996})}\BibitemShut {NoStop}%
\bibitem [{\citenamefont {Niemeyer}\ \emph {et~al.}(1999)\citenamefont
  {Niemeyer}, \citenamefont {Freericks},\ and\ \citenamefont
  {Monien}}]{niemeyer_99}%
  \BibitemOpen
  \bibfield  {author} {\bibinfo {author} {\bibfnamefont {M.}~\bibnamefont
  {Niemeyer}}, \bibinfo {author} {\bibfnamefont {J.~K.}\ \bibnamefont
  {Freericks}}, \ and\ \bibinfo {author} {\bibfnamefont {H.}~\bibnamefont
  {Monien}},\ }\bibfield  {title} {\enquote {\bibinfo {title} {Strong-coupling
  perturbation theory for the two-dimensional {B}ose-{H}ubbard model in a
  magnetic field},}\ }\href {\doibase 10.1103/PhysRevB.60.2357} {\bibfield
  {journal} {\bibinfo  {journal} {Phys. Rev. B}\ }\textbf {\bibinfo {volume}
  {60}},\ \bibinfo {pages} {2357--2362} (\bibinfo {year} {1999})}\BibitemShut
  {NoStop}%
\bibitem [{\citenamefont {Freericks}\ \emph {et~al.}(2009)\citenamefont
  {Freericks}, \citenamefont {Krishnamurthy}, \citenamefont {Kato},
  \citenamefont {Kawashima},\ and\ \citenamefont {Trivedi}}]{freericks_09}%
  \BibitemOpen
  \bibfield  {author} {\bibinfo {author} {\bibfnamefont {J.~K.}\ \bibnamefont
  {Freericks}}, \bibinfo {author} {\bibfnamefont {H.~R.}\ \bibnamefont
  {Krishnamurthy}}, \bibinfo {author} {\bibfnamefont {Yasuyuki}\ \bibnamefont
  {Kato}}, \bibinfo {author} {\bibfnamefont {Naoki}\ \bibnamefont {Kawashima}},
  \ and\ \bibinfo {author} {\bibfnamefont {Nandini}\ \bibnamefont {Trivedi}},\
  }\bibfield  {title} {\enquote {\bibinfo {title} {Strong-coupling expansion
  for the momentum distribution of the bose-hubbard model with benchmarking
  against exact numerical results},}\ }\href {\doibase
  10.1103/PhysRevA.79.053631} {\bibfield  {journal} {\bibinfo  {journal} {Phys.
  Rev. A}\ }\textbf {\bibinfo {volume} {79}},\ \bibinfo {pages} {053631}
  (\bibinfo {year} {2009})}\BibitemShut {NoStop}%
\bibitem [{\citenamefont {Wang}\ \emph {et~al.}(2018)\citenamefont {Wang},
  \citenamefont {Zhang}, \citenamefont {Hou}, \citenamefont {Eggert},\ and\
  \citenamefont {Pelster}}]{wang_18}%
  \BibitemOpen
  \bibfield  {author} {\bibinfo {author} {\bibfnamefont {T.}~\bibnamefont
  {Wang}}, \bibinfo {author} {\bibfnamefont {X.~F.}\ \bibnamefont {Zhang}},
  \bibinfo {author} {\bibfnamefont {C.~F.}\ \bibnamefont {Hou}}, \bibinfo
  {author} {\bibfnamefont {S.}~\bibnamefont {Eggert}}, \ and\ \bibinfo {author}
  {\bibfnamefont {A.}~\bibnamefont {Pelster}},\ }\bibfield  {title} {\enquote
  {\bibinfo {title} {High-order strong-coupling expansion for the
  {B}ose-{H}ubbard model},}\ }\href {https://arxiv.org/abs/1801.01862}
  {\bibfield  {journal} {\bibinfo  {journal} {arXiv:1801.01862}\ } (\bibinfo
  {year} {2018})}\BibitemShut {NoStop}%
\bibitem [{\citenamefont {Wessel}\ \emph {et~al.}(2004)\citenamefont {Wessel},
  \citenamefont {Alet}, \citenamefont {Troyer},\ and\ \citenamefont
  {Batrouni}}]{wessel_04}%
  \BibitemOpen
  \bibfield  {author} {\bibinfo {author} {\bibfnamefont {S.}~\bibnamefont
  {Wessel}}, \bibinfo {author} {\bibfnamefont {F.}~\bibnamefont {Alet}},
  \bibinfo {author} {\bibfnamefont {M.}~\bibnamefont {Troyer}}, \ and\ \bibinfo
  {author} {\bibfnamefont {G.G.}\ \bibnamefont {Batrouni}},\ }\bibfield
  {title} {\enquote {\bibinfo {title} {Quantum {M}onte {C}arlo simulations of
  confined bosonic atoms in optical lattices},}\ }\href {\doibase
  10.1103/PhysRevA.70.053615} {\bibfield  {journal} {\bibinfo  {journal} {Phys.
  Rev. A}\ }\textbf {\bibinfo {volume} {70}},\ \bibinfo {pages} {053615}
  (\bibinfo {year} {2004})}\BibitemShut {NoStop}%
\bibitem [{\citenamefont {Peotta}\ \emph {et~al.}(2014)\citenamefont {Peotta},
  \citenamefont {Chien},\ and\ \citenamefont {Di~Ventra}}]{peotta_14}%
  \BibitemOpen
  \bibfield  {author} {\bibinfo {author} {\bibfnamefont {S.}~\bibnamefont
  {Peotta}}, \bibinfo {author} {\bibfnamefont {C.-C.}\ \bibnamefont {Chien}}, \
  and\ \bibinfo {author} {\bibfnamefont {M.}~\bibnamefont {Di~Ventra}},\
  }\bibfield  {title} {\enquote {\bibinfo {title} {Phase-induced transport in
  atomic gases: From superfluid to {M}ott insulator},}\ }\href {\doibase
  10.1103/PhysRevA.90.053615} {\bibfield  {journal} {\bibinfo  {journal} {Phys.
  Rev. A}\ }\textbf {\bibinfo {volume} {90}},\ \bibinfo {pages} {053615}
  (\bibinfo {year} {2014})}\BibitemShut {NoStop}%
\bibitem [{\citenamefont {Hafezi}\ \emph {et~al.}(2007)\citenamefont {Hafezi},
  \citenamefont {S\o{}rensen}, \citenamefont {Demler},\ and\ \citenamefont
  {Lukin}}]{hafezi_07}%
  \BibitemOpen
  \bibfield  {author} {\bibinfo {author} {\bibfnamefont {M.}~\bibnamefont
  {Hafezi}}, \bibinfo {author} {\bibfnamefont {A.~S.}\ \bibnamefont
  {S\o{}rensen}}, \bibinfo {author} {\bibfnamefont {E.}~\bibnamefont {Demler}},
  \ and\ \bibinfo {author} {\bibfnamefont {M.~D.}\ \bibnamefont {Lukin}},\
  }\bibfield  {title} {\enquote {\bibinfo {title} {Fractional quantum {H}all
  effect in optical lattices},}\ }\href {\doibase 10.1103/PhysRevA.76.023613}
  {\bibfield  {journal} {\bibinfo  {journal} {Phys. Rev. A}\ }\textbf {\bibinfo
  {volume} {76}},\ \bibinfo {pages} {023613} (\bibinfo {year}
  {2007})}\BibitemShut {NoStop}%
\bibitem [{\citenamefont {Umucal\ifmmode \imath \else~\i \fi{}lar}\ and\
  \citenamefont {Oktel}(2007)}]{umucalilar_07}%
  \BibitemOpen
  \bibfield  {author} {\bibinfo {author} {\bibfnamefont {R.~O.}\ \bibnamefont
  {Umucal\ifmmode \imath \else~\i \fi{}lar}}\ and\ \bibinfo {author}
  {\bibfnamefont {M.~\"O.}\ \bibnamefont {Oktel}},\ }\bibfield  {title}
  {\enquote {\bibinfo {title} {Phase boundary of the boson mott insulator in a
  rotating optical lattice},}\ }\href {\doibase 10.1103/PhysRevA.76.055601}
  {\bibfield  {journal} {\bibinfo  {journal} {Phys. Rev. A}\ }\textbf {\bibinfo
  {volume} {76}},\ \bibinfo {pages} {055601} (\bibinfo {year}
  {2007})}\BibitemShut {NoStop}%
\bibitem [{\citenamefont {Palmer}\ \emph {et~al.}(2008)\citenamefont {Palmer},
  \citenamefont {Klein},\ and\ \citenamefont {Jaksch}}]{palmer_08}%
  \BibitemOpen
  \bibfield  {author} {\bibinfo {author} {\bibfnamefont {R.~N.}\ \bibnamefont
  {Palmer}}, \bibinfo {author} {\bibfnamefont {A.}~\bibnamefont {Klein}}, \
  and\ \bibinfo {author} {\bibfnamefont {D.}~\bibnamefont {Jaksch}},\
  }\bibfield  {title} {\enquote {\bibinfo {title} {Optical lattice quantum
  {H}all effect},}\ }\href {\doibase 10.1103/PhysRevA.78.013609} {\bibfield
  {journal} {\bibinfo  {journal} {Phys. Rev. A}\ }\textbf {\bibinfo {volume}
  {78}},\ \bibinfo {pages} {013609} (\bibinfo {year} {2008})}\BibitemShut
  {NoStop}%
\bibitem [{\citenamefont {Umucalilar}\ and\ \citenamefont
  {Mueller}(2010)}]{umucalilar_10}%
  \BibitemOpen
  \bibfield  {author} {\bibinfo {author} {\bibfnamefont {R.~O.}\ \bibnamefont
  {Umucalilar}}\ and\ \bibinfo {author} {\bibfnamefont {E.~J.}\ \bibnamefont
  {Mueller}},\ }\bibfield  {title} {\enquote {\bibinfo {title} {Fractional
  quantum {H}all states in the vicinity of {M}ott plateaus},}\ }\href {\doibase
  10.1103/PhysRevA.81.053628} {\bibfield  {journal} {\bibinfo  {journal} {Phys.
  Rev. A}\ }\textbf {\bibinfo {volume} {81}},\ \bibinfo {pages} {053628}
  (\bibinfo {year} {2010})}\BibitemShut {NoStop}%
\bibitem [{\citenamefont {Natu}\ \emph {et~al.}(2016)\citenamefont {Natu},
  \citenamefont {Mueller},\ and\ \citenamefont {Das~Sarma}}]{natu_16}%
  \BibitemOpen
  \bibfield  {author} {\bibinfo {author} {\bibfnamefont {S.~S.}\ \bibnamefont
  {Natu}}, \bibinfo {author} {\bibfnamefont {E.~J.}\ \bibnamefont {Mueller}}, \
  and\ \bibinfo {author} {\bibfnamefont {S.}~\bibnamefont {Das~Sarma}},\
  }\bibfield  {title} {\enquote {\bibinfo {title} {Competing ground states of
  strongly correlated bosons in the {H}arper-{H}ofstadter-{M}ott model},}\
  }\href {\doibase 10.1103/PhysRevA.93.063610} {\bibfield  {journal} {\bibinfo
  {journal} {Phys. Rev. A}\ }\textbf {\bibinfo {volume} {93}},\ \bibinfo
  {pages} {063610} (\bibinfo {year} {2016})}\BibitemShut {NoStop}%
\bibitem [{\citenamefont {H\"ugel}\ \emph {et~al.}(2017)\citenamefont
  {H\"ugel}, \citenamefont {Strand}, \citenamefont {Werner},\ and\
  \citenamefont {Pollet}}]{hugel_17}%
  \BibitemOpen
  \bibfield  {author} {\bibinfo {author} {\bibfnamefont {D.}~\bibnamefont
  {H\"ugel}}, \bibinfo {author} {\bibfnamefont {H.~U.~R.}\ \bibnamefont
  {Strand}}, \bibinfo {author} {\bibfnamefont {P.}~\bibnamefont {Werner}}, \
  and\ \bibinfo {author} {\bibfnamefont {L.}~\bibnamefont {Pollet}},\
  }\bibfield  {title} {\enquote {\bibinfo {title} {Anisotropic
  {H}arper-{H}ofstadter-{M}ott model: Competition between condensation and
  magnetic fields},}\ }\href {\doibase 10.1103/PhysRevB.96.054431} {\bibfield
  {journal} {\bibinfo  {journal} {Phys. Rev. B}\ }\textbf {\bibinfo {volume}
  {96}},\ \bibinfo {pages} {054431} (\bibinfo {year} {2017})}\BibitemShut
  {NoStop}%
\bibitem [{\citenamefont {Kuno}\ \emph {et~al.}(2017)\citenamefont {Kuno},
  \citenamefont {Shimizu},\ and\ \citenamefont {Ichinose}}]{kuno_17}%
  \BibitemOpen
  \bibfield  {author} {\bibinfo {author} {\bibfnamefont {Y.}~\bibnamefont
  {Kuno}}, \bibinfo {author} {\bibfnamefont {K.}~\bibnamefont {Shimizu}}, \
  and\ \bibinfo {author} {\bibfnamefont {I.}~\bibnamefont {Ichinose}},\
  }\bibfield  {title} {\enquote {\bibinfo {title} {Bosonic analogs of the
  fractional quantum {H}all state in the vicinity of mott states},}\ }\href
  {\doibase 10.1103/PhysRevA.95.013607} {\bibfield  {journal} {\bibinfo
  {journal} {Phys. Rev. A}\ }\textbf {\bibinfo {volume} {95}},\ \bibinfo
  {pages} {013607} (\bibinfo {year} {2017})}\BibitemShut {NoStop}%
\bibitem [{\citenamefont {Gerster}\ \emph {et~al.}(2017)\citenamefont
  {Gerster}, \citenamefont {Rizzi}, \citenamefont {Silvi}, \citenamefont
  {Dalmonte},\ and\ \citenamefont {Montangero}}]{gerster_17}%
  \BibitemOpen
  \bibfield  {author} {\bibinfo {author} {\bibfnamefont {M.}~\bibnamefont
  {Gerster}}, \bibinfo {author} {\bibfnamefont {M.}~\bibnamefont {Rizzi}},
  \bibinfo {author} {\bibfnamefont {P.}~\bibnamefont {Silvi}}, \bibinfo
  {author} {\bibfnamefont {M.}~\bibnamefont {Dalmonte}}, \ and\ \bibinfo
  {author} {\bibfnamefont {S.}~\bibnamefont {Montangero}},\ }\bibfield  {title}
  {\enquote {\bibinfo {title} {Fractional quantum hall effect in the
  interacting hofstadter model via tensor networks},}\ }\href {\doibase
  10.1103/PhysRevB.96.195123} {\bibfield  {journal} {\bibinfo  {journal} {Phys.
  Rev. B}\ }\textbf {\bibinfo {volume} {96}},\ \bibinfo {pages} {195123}
  (\bibinfo {year} {2017})}\BibitemShut {NoStop}%
\bibitem [{\citenamefont {Bai}\ \emph {et~al.}(2018)\citenamefont {Bai},
  \citenamefont {Bandyopadhyay}, \citenamefont {Pal}, \citenamefont {Suthar},\
  and\ \citenamefont {Angom}}]{bai_18}%
  \BibitemOpen
  \bibfield  {author} {\bibinfo {author} {\bibfnamefont {Rukmani}\ \bibnamefont
  {Bai}}, \bibinfo {author} {\bibfnamefont {Soumik}\ \bibnamefont
  {Bandyopadhyay}}, \bibinfo {author} {\bibfnamefont {Sukla}\ \bibnamefont
  {Pal}}, \bibinfo {author} {\bibfnamefont {K.}~\bibnamefont {Suthar}}, \ and\
  \bibinfo {author} {\bibfnamefont {D.}~\bibnamefont {Angom}},\ }\bibfield
  {title} {\enquote {\bibinfo {title} {Bosonic quantum hall states in
  single-layer two-dimensional optical lattices},}\ }\href {\doibase
  10.1103/PhysRevA.98.023606} {\bibfield  {journal} {\bibinfo  {journal} {Phys.
  Rev. A}\ }\textbf {\bibinfo {volume} {98}},\ \bibinfo {pages} {023606}
  (\bibinfo {year} {2018})}\BibitemShut {NoStop}%
\bibitem [{\citenamefont {He}\ \emph {et~al.}(2017)\citenamefont {He},
  \citenamefont {Grusdt}, \citenamefont {Kaufman}, \citenamefont {Greiner},\
  and\ \citenamefont {Vishwanath}}]{he_17}%
  \BibitemOpen
  \bibfield  {author} {\bibinfo {author} {\bibfnamefont {Y.-C.}\ \bibnamefont
  {He}}, \bibinfo {author} {\bibfnamefont {F.}~\bibnamefont {Grusdt}}, \bibinfo
  {author} {\bibfnamefont {A.}~\bibnamefont {Kaufman}}, \bibinfo {author}
  {\bibfnamefont {M.}~\bibnamefont {Greiner}}, \ and\ \bibinfo {author}
  {\bibfnamefont {A.}~\bibnamefont {Vishwanath}},\ }\bibfield  {title}
  {\enquote {\bibinfo {title} {Realizing and adiabatically preparing bosonic
  integer and fractional quantum {H}all states in optical lattices},}\ }\href
  {\doibase 10.1103/PhysRevB.96.201103} {\bibfield  {journal} {\bibinfo
  {journal} {Phys. Rev. B}\ }\textbf {\bibinfo {volume} {96}},\ \bibinfo
  {pages} {201103} (\bibinfo {year} {2017})}\BibitemShut {NoStop}%
\bibitem [{\citenamefont {Elliott}\ and\ \citenamefont
  {Johnson}(2016)}]{elliott_16}%
  \BibitemOpen
  \bibfield  {author} {\bibinfo {author} {\bibfnamefont {T.~J.}\ \bibnamefont
  {Elliott}}\ and\ \bibinfo {author} {\bibfnamefont {T.~H.}\ \bibnamefont
  {Johnson}},\ }\bibfield  {title} {\enquote {\bibinfo {title} {Nondestructive
  probing of means, variances, and correlations of ultracold-atomic-system
  densities via qubit impurities},}\ }\href {\doibase
  10.1103/PhysRevA.93.043612} {\bibfield  {journal} {\bibinfo  {journal} {Phys.
  Rev. A}\ }\textbf {\bibinfo {volume} {93}},\ \bibinfo {pages} {043612}
  (\bibinfo {year} {2016})}\BibitemShut {NoStop}%
\bibitem [{\citenamefont {Streif}\ \emph {et~al.}(2016)\citenamefont {Streif},
  \citenamefont {Buchleitner}, \citenamefont {Jaksch},\ and\ \citenamefont
  {Mur-Petit}}]{streif_16}%
  \BibitemOpen
  \bibfield  {author} {\bibinfo {author} {\bibfnamefont {M.}~\bibnamefont
  {Streif}}, \bibinfo {author} {\bibfnamefont {A.}~\bibnamefont {Buchleitner}},
  \bibinfo {author} {\bibfnamefont {D.}~\bibnamefont {Jaksch}}, \ and\ \bibinfo
  {author} {\bibfnamefont {J.}~\bibnamefont {Mur-Petit}},\ }\bibfield  {title}
  {\enquote {\bibinfo {title} {Measuring correlations of cold-atom systems
  using multiple quantum probes},}\ }\href {\doibase
  10.1103/PhysRevA.94.053634} {\bibfield  {journal} {\bibinfo  {journal} {Phys.
  Rev. A}\ }\textbf {\bibinfo {volume} {94}},\ \bibinfo {pages} {053634}
  (\bibinfo {year} {2016})}\BibitemShut {NoStop}%
\bibitem [{\citenamefont {Aidelsburger}\ \emph {et~al.}(2011)\citenamefont
  {Aidelsburger}, \citenamefont {Atala}, \citenamefont {Nascimb\`ene},
  \citenamefont {Trotzky}, \citenamefont {Chen},\ and\ \citenamefont
  {Bloch}}]{aidelsburger_11}%
  \BibitemOpen
  \bibfield  {author} {\bibinfo {author} {\bibfnamefont {M.}~\bibnamefont
  {Aidelsburger}}, \bibinfo {author} {\bibfnamefont {M.}~\bibnamefont {Atala}},
  \bibinfo {author} {\bibfnamefont {S.}~\bibnamefont {Nascimb\`ene}}, \bibinfo
  {author} {\bibfnamefont {S.}~\bibnamefont {Trotzky}}, \bibinfo {author}
  {\bibfnamefont {Y.-A.}\ \bibnamefont {Chen}}, \ and\ \bibinfo {author}
  {\bibfnamefont {I.}~\bibnamefont {Bloch}},\ }\bibfield  {title} {\enquote
  {\bibinfo {title} {Experimental realization of strong effective magnetic
  fields in an optical lattice},}\ }\href {\doibase
  10.1103/PhysRevLett.107.255301} {\bibfield  {journal} {\bibinfo  {journal}
  {Phys. Rev. Lett.}\ }\textbf {\bibinfo {volume} {107}},\ \bibinfo {pages}
  {255301} (\bibinfo {year} {2011})}\BibitemShut {NoStop}%
\bibitem [{\citenamefont {Oktel}\ \emph {et~al.}(2007)\citenamefont {Oktel},
  \citenamefont {Ni\ifmmode \mbox{\c{t}}\else \c{t}\fi{} \ifmmode~\u{a}\else
  \u{a}\fi{}},\ and\ \citenamefont {Tanatar}}]{oktel_07}%
  \BibitemOpen
  \bibfield  {author} {\bibinfo {author} {\bibfnamefont {M.~\"O.}\ \bibnamefont
  {Oktel}}, \bibinfo {author} {\bibfnamefont {M.}~\bibnamefont {Ni\ifmmode
  \mbox{\c{t}}\else \c{t}\fi{} \ifmmode~\u{a}\else \u{a}\fi{}}}, \ and\
  \bibinfo {author} {\bibfnamefont {B.}~\bibnamefont {Tanatar}},\ }\bibfield
  {title} {\enquote {\bibinfo {title} {Mean-field theory for {B}ose-{H}ubbard
  model under a magnetic field},}\ }\href {\doibase 10.1103/PhysRevB.75.045133}
  {\bibfield  {journal} {\bibinfo  {journal} {Phys. Rev. B}\ }\textbf {\bibinfo
  {volume} {75}},\ \bibinfo {pages} {045133} (\bibinfo {year}
  {2007})}\BibitemShut {NoStop}%
\bibitem [{\citenamefont {Sheshadri}\ \emph {et~al.}(1993)\citenamefont
  {Sheshadri}, \citenamefont {Krishnamurthy}, \citenamefont {Pandit},\ and\
  \citenamefont {Ramakrishnan}}]{sheshadri_93}%
  \BibitemOpen
  \bibfield  {author} {\bibinfo {author} {\bibfnamefont {K.}~\bibnamefont
  {Sheshadri}}, \bibinfo {author} {\bibfnamefont {H.~R.}\ \bibnamefont
  {Krishnamurthy}}, \bibinfo {author} {\bibfnamefont {R.}~\bibnamefont
  {Pandit}}, \ and\ \bibinfo {author} {\bibfnamefont {T.~V.}\ \bibnamefont
  {Ramakrishnan}},\ }\bibfield  {title} {\enquote {\bibinfo {title} {Superfluid
  and insulating phases in an interacting-boson model: Mean-field theory and
  the {RPA}},}\ }\href {\doibase https://doi.org/10.1209/0295-5075/22/4/004}
  {\bibfield  {journal} {\bibinfo  {journal} {EPL}\ }\textbf {\bibinfo {volume}
  {22}},\ \bibinfo {pages} {257} (\bibinfo {year} {1993})}\BibitemShut
  {NoStop}%
\bibitem [{\citenamefont {L\"uhmann}(2013)}]{luhmann_13}%
  \BibitemOpen
  \bibfield  {author} {\bibinfo {author} {\bibfnamefont {D.-S.}\ \bibnamefont
  {L\"uhmann}},\ }\bibfield  {title} {\enquote {\bibinfo {title} {Cluster
  {G}utzwiller method for bosonic lattice systems},}\ }\href {\doibase
  10.1103/PhysRevA.87.043619} {\bibfield  {journal} {\bibinfo  {journal} {Phys.
  Rev. A}\ }\textbf {\bibinfo {volume} {87}},\ \bibinfo {pages} {043619}
  (\bibinfo {year} {2013})}\BibitemShut {NoStop}%
\bibitem [{\citenamefont {Barrett}\ \emph {et~al.}(1994)\citenamefont
  {Barrett}, \citenamefont {Berry}, \citenamefont {Chan}, \citenamefont
  {Demmel}, \citenamefont {Donato}, \citenamefont {Dongarra}, \citenamefont
  {Eijkhout}, \citenamefont {Pozo}, \citenamefont {Romine},\ and\ \citenamefont
  {van~der Vorst}}]{barrett_94}%
  \BibitemOpen
  \bibfield  {author} {\bibinfo {author} {\bibfnamefont {R.}~\bibnamefont
  {Barrett}}, \bibinfo {author} {\bibfnamefont {M.}~\bibnamefont {Berry}},
  \bibinfo {author} {\bibfnamefont {T.}~\bibnamefont {Chan}}, \bibinfo {author}
  {\bibfnamefont {J.}~\bibnamefont {Demmel}}, \bibinfo {author} {\bibfnamefont
  {J.}~\bibnamefont {Donato}}, \bibinfo {author} {\bibfnamefont
  {J.}~\bibnamefont {Dongarra}}, \bibinfo {author} {\bibfnamefont
  {V.}~\bibnamefont {Eijkhout}}, \bibinfo {author} {\bibfnamefont
  {R.}~\bibnamefont {Pozo}}, \bibinfo {author} {\bibfnamefont {C.}~\bibnamefont
  {Romine}}, \ and\ \bibinfo {author} {\bibfnamefont {H.}~\bibnamefont {van~der
  Vorst}},\ }\href {\doibase 10.1137/1.9781611971538} {\emph {\bibinfo {title}
  {Templates for the Solution of Linear Systems: Building Blocks for Iterative
  Methods}}}\ (\bibinfo  {publisher} {SIAM},\ \bibinfo {year}
  {1994})\BibitemShut {NoStop}%
\end{thebibliography}%

\end{document}